\documentclass[
  twoside,
  twocolumn,
  amsmath,
  amssymb,
  aps,
  prstab,
  longbibliography,
  nofootinbib,
  floatfix,
]{revtex4-1}

\RequirePackage{microtype}
\RequirePackage{graphicx}
\graphicspath{{./}{./figures/}}

\RequirePackage{dcolumn}
\RequirePackage{bm}
\RequirePackage{liecolon}
\RequirePackage{units}
\RequirePackage{mathrsfs}
\DeclareMathAlphabet{\mathpzc}{OT1}{pzc}{m}{it}

\RequirePackage{suffix}
\newcommand*{\figurerefname}{figure}
\newcommand*{\Figurerefname}{Figure}
\newcommand*{\figuresrefname}{figures}
\newcommand*{\Figuresrefname}{Figures}
\newcommand*{\fref}[2][]{{%
  \ifthenelse{\isempty{#1}}{\figurerefname}{%
   \ifthenelse{\equal{#1}{s}}{\figuresrefname}{%
    \ifthenelse{\equal{#1}{c}}{\Figurerefname}{%
     \ifthenelse{\equal{#1}{cs}\OR\equal{#1}{sc}}{\Figuresrefname}{%
      \typeout{cross-ref error -- fig:#2}\figurerefname}}}}~\ref{fig:#2}}}
\WithSuffix\def\fref*#1{\ref{fig:#1}}
\newcommand*{\tablerefname}{table}
\newcommand*{\Tablerefname}{Table}
\newcommand*{\tablesrefname}{tables}
\newcommand*{\Tablesrefname}{Tables}
\newcommand*{\tref}[2][]{{%
  \ifthenelse{\isempty{#1}}{\tablerefname}{%
   \ifthenelse{\equal{#1}{s}}{\tablesrefname}{%
    \ifthenelse{\equal{#1}{c}}{\Tablerefname}{%
     \ifthenelse{\equal{#1}{cs}\OR\equal{#1}{sc}}{\Tablesrefname}{%
      \typeout{cross-ref error -- tbl:#2}\tablerefname}}}}~\ref{tbl:#2}}}
\WithSuffix\def\tref*#1{\ref{tbl:#1}}
\newcommand*{\equationrefname}{}
\newcommand*{\Equationrefname}{Equation~}
\newcommand*{\equationsrefname}{}
\newcommand*{\Equationsrefname}{Equations~}
\renewcommand*{\eqref}[2][]{{%
  \ifthenelse{\isempty{#1}}{\equationrefname}{%
   \ifthenelse{\equal{#1}{s}}{\equationsrefname}{%
    \ifthenelse{\equal{#1}{c}}{\Equationrefname}{%
     \ifthenelse{\equal{#1}{cs}\OR\equal{#1}{sc}}{\Equationsrefname}{%
      \typeout{cross-ref error -- eq:#2}\equationrefname}}}}(\ref{eq:#2})}}
\WithSuffix\def\eqref*#1{(\ref{eq:#1})}

\newcommand{\ud}{\mathop{}\!\mathrm{d}}
\newcommand{\dd}[3][]{\ifthenelse{\isempty{#1}}%
  {\ud{#2}/\ud{#3}}%
  {\ud^{#1}{#2}/\ud{#3}^{#1}}}
\newcommand{\Dd}[3][]{\ifthenelse{\isempty{#1}}%
  {\frac{\ud{#2}}{\ud{#3}}}%
  {\frac{\ud^{#1}{#2}}{\ud{#3}^{#1}}}}
\newcommand{\dt}[2][]{\ifthenelse{\isempty{#1}}%
  {\ud{#2}/\ud{t}}%
  {\ud^{#1}{#2}/\ud{t}^{#1}}}
\newcommand{\Dt}[2][]{\ifthenelse{\isempty{#1}}%
  {\frac{\ud{#2}}{\ud{t}}}%
  {\frac{\ud^{#1}{#2}}{\ud{t}^{#1}}}}
\newcommand{\ptdd}[3][]{\ifthenelse{\isempty{#1}}%
  {\partial{#2}/\partial{#3}}%
  {\partial^{#1}{#2}/\partial{#3}^{#1}}}
\newcommand{\ptDd}[3][]{\ifthenelse{\isempty{#1}}%
  {\frac{\partial{#2}}{\partial{#3}}}%
  {\frac{\partial^{#1}{#2}}{\partial{#3}^{#1}}}}
\DeclareMathOperator{\grad}{\nabla}

\DeclareMathOperator{\curl}{\nabla\times}
\newcommand{\Hpc}{\mathpzc{H}_\text{pc}}
\newcommand{\Hf}{\mathpzc{H}_\text{f}}
\newcommand{\Mpc}{\mathscr{M}_\text{pc}}
\newcommand{\Mf}{\mathscr{M}_\text{f}}
\newcommand{\Qell}{\mathcal{Q}_\ell}
\newcommand{\Pell}{\mathcal{P}_\ell}

\begin{document}

\preprint{}

\title{Symplectic Modeling of Beam Loading in Electromagnetic Cavities}

\author{Dan T. Abell}
  \email{dabell@radiasoft.net}
  \homepage{www.radiasoft.net}
\author{Nathan M. Cook}
\author{Stephen D. Webb}

\affiliation{RadiaSoft, LLC, 1348 Redwood Ave., Boulder, CO 80304}
 \altaffiliation[]{}

\date{\today}

\begin{abstract}
Simulating beam loading in radiofrequency accelerating structures is critical for understanding higher-order mode effects on beam dynamics, such as beam break-up instability in energy recovery linacs. Full wave simulations of beam loading in radiofrequency structures are computationally expensive, while reduced models can ignore essential physics and can be difficult to generalize. We present a self-consistent algorithm derived from the least-action principle which can model an arbitrary number of cavity eigenmodes and with a generic beam distribution.
\end{abstract}


\maketitle

\section{Introduction}

The interactions of charged-particle beams and electromagnetic fields
drive a variety of applications and phenomena in accelerator systems.
They range from klystrons to beam-loading in radiofrequency (RF) cavities
to high-order-mode instabilities in energy-recovery linacs.
Accurate, self-consistent modeling of these interactions is an important
and challenging problem for many of these applications.

Self-consistent simulations using approaches such as
finite-difference time-domain (FDTD) particle-in-cell (PIC) algorithms%
  ~\cite{birdsall_langdon:85}
are computationally demanding and suffer from a variety of numerical
artifacts---for example, the numerical \v{C}erenkov instability%
  ~\cite{godfrey:74},
or dispersive errors induced by the mesh%
  ~\cite{Trefethen:1982:GrpVelFDSchemes,
         taflove_hagness}%
---that can muddy the results.
Detailed simulations of an electron beam passing through an RF cavity
must resolve the current (i.e. the electron beam), and hence require
cell dimensions that are small compared to the beam.  Resolving the beam
across the entire complex geometry can therefore require billions of cells,
while the actual source terms are isolated to a mere few thousands of cells.
As a consequence, load balancing becomes a serious concern for multi-core
simulations.

The discrepancy of spatial scales is not the only challenge faced by
FDTD-PIC algorithms.
One must also initialize the fields on the numerical mesh in a manner
that satisfies the appropriate numerical dispersion relation%
  ~\cite{Trefethen:1982:GrpVelFDSchemes,
         Vay:2010:ModelLaserLorentzBoost,
         Godfrey:2012:NumerStabPIC}.
In addition, the presence of multiple bunches means that one must address
the discrepancy of temporal scales that effectively prevents using such
algorithms to study the effect of high-order modes on beam dynamics.

To bridge multiple scales, one often resorts to reduced models, as, for
example, when studying
beam loading (\texttt{matbbu}~\cite{beard_merminga_yunn:03})
or
beam instability (\texttt{bi}~\cite{bazarov:03}).
Such models typically treat
the RF cavity as a thin lens, using reduced forms of the Shockley-Ramo
theorem%
  ~\cite{shockley:38,ramo:39}
to compute the energy transfer; and they can advance the field phases
analytically once the beam passes.  Reduced models are computationally
much more efficient than full simulations.  They also simplify the
diagnostics, because the energy in each cavity mode is a dynamical
variable and requires no additional computation to be extracted from
the simulation.  However, they can fail to demonstrate complicated
phase-space structures that result from beam-beam disruption in a collider
(as in the eRHIC energy recovery linac-ring design for an electron-ion
collider%
  ~\cite{Hao:2013:MitigatKinkInstab,
         Hao:2014:BmBmStudyERLeRHIC})
or from a large energy spread (as results from advanced tapering schemes
for free-electron lasers).  The approximations made in reduced models
frequently do not fail gracefully, and they are difficult to expand to
next-leading-order, which can limit their versatility.

What we require is an algorithm that has the simplicity and physically
intuitive feel of a reduced model, while being extensible for
self-consistent simulations. In this paper, we present such an algorithm
based on a symplectic map approach to electromagnetic particle-in-mode
algorithms%
  ~\cite{webb_etal:2016c}.
For this approach we use the cavity eigenmodes as our orthonormal basis
for the electromagnetic field. We consider only the coupling of $j_z$ to $A_z$,
neglecting the transverse currents.
And the modes themselves we compute using either an analytic model or
interpolation of numerical data~\cite{abell:06}.
This approach leads to a fast, extensible model for beam loading with
arbitrary numbers of modes and cavity geometries.

\section{The Rationale for a Spectral Time-based Algorithm}

Self-consistent updates of the Maxwell equations require that the fields
obey the boundary conditions.  For FDTD-type algorithms---the canonical
example being the staggered-Yee scheme%
  ~\cite{yee:66}%
---this is not difficult: the basis used to decompose the fields
is local in space, and it is easy to compute reflections off boundaries,
transient effects, etc.
FDTD algorithms are popular also because they
can handle complex boundaries, and the electromagnetic part of the
algorithm requires only local information to update the fields.
However, these algorithms must resolve the smallest features in a simulation,
typically the beam itself, and they have inherent dispersive errors that
make accurate simulation of beam loading a challenge.
Artificial instabilities such as numerical \v{C}erenkov add to the
challenge, thus overall making the FDTD approach unsuitable for studying
beam loading in most structures.

One possible solution is to use a spectral algorithm that does a global
decomposition of the fields as a superposition of the cavity eigenmodes.
This is similar in spirit to the Condon method for computing wakefields%
  ~\cite{Condon:1941:ForcedOscCavRes,
       Wilson:1989:IntroWakefields,
       Chao:1993:PhysCBI}.
Each mode frequency is known exactly%
  \footnote{Or to within numerical tolerances, if the fields are
            solved numerically using a frequency domain solver.},
meaning that in the absence of a source term it is possible to evolve
the fields exactly. Adding the source term---the beam---is a convolution
integral of the source currents and charges with each field eigenmode.
The beam itself takes up a small volume of the simulation domain.
Using the field eigenmodes removes the longest length scale from the
simulation, dramatically reducing the computational resources required.
It also greatly simplifies source deposition in an electromagnetic
simulation, as compared to deposition for FDTD, which can encounter
load balancing issues that impede parallel scaling%
  \footnote{It is worth noting that because spectral algorithms are
    global instead of local algorithms, they require an \texttt{MPI\_AllReduce}
    to consolidate the source terms. This prevents spectral algorithms from
    scaling indefinitely, and may require specialized algorithms which reduce
    the need for many-core simulations at the expense of generality.}.
    This can make the spectral approach competitive for performance with FDTD
    algorithms. Because beam loading requires exact frequency information and
    field maps to compute properly, and the sources are located on a small
    fraction of the simulation domain which causes severe load balancing
    issues for FDTD particle-in-cell simulations, spectral algorithms are
    ideal for self-consistent beam loading simulations.

Conventional accelerator tracking codes almost universally use distance along the design orbit of a ring or linac as the independent variable ($s$-based tracking). When including collective effects, a code must address the issue of simultaneity in the Poisson equation: that leads to codes either adopting time as the independent variable ($t$-based tracking), or using various tricks to make the particle times simultaneous. Single particle integration through standing wave structures, such as radiofrequency cavities, can be accomplished so long as the self-consistent evolution of the fields is neglected.

In a waveguide type structure, which has translational symmetry in the longitudinal direction, $s$- or $t$-based tracking can be used. For systems with $s$-varying transverse geometries, such as RF cavities or traveling wave tubes, we must use $t$-based tracking. It is easiest to understand this by looking at how an ultra-relativistic beam ($\beta \approx 1$) would affect an $s$- versus $t$-based spectral representation. The fields from the bunch radiate purely transversely, before encountering the cavity surface. If that surface has any tilt to it, the fields reflect off the cavity surface and back into the cavity. In an $s$-based approach, this would produce a local-in-$s$ increase in the individual cavity eigenmode strengths on the surface, creating surface currents. These surface currents are not self-consistent with any single eigenmode, so the fields would re-radiate. An $s$-based eigenmode representation has a scattering term between multiple eigenmodes. This is not the case for a $t$-based approach, which changes the eigenmode amplitudes and their associated surface currents by a multiplicative constant, but which are still self-consistent for a single eigenmode.

Simply, an $s$-based algorithm with $s$-varying geometry has scattering terms
between the eigenmodes, while a $t$-based algorithm would not. For the problem
of a standing wave rf cavity, this means that a $t$-based algorithm is simpler.
For a waveguide with no longitudinal variation in the boundaries, either
$s$- or $t$-based algorithms can be used.

\section{Self-Consistent Beam Loading Algorithm}

\subsection{Beam Low Lagrangian}

We begin with the \emph{Low Lagrangian}%
  ~\cite{low:58},
devised by Low in 1958 to provide a variational formulation for describing
a non-relativistic ionized gas in an electromagnetic field.
It has been used, for example, to study nonlinear waves in plasmas%
  ~\cite{Dewar:1972:LangrangTheorNLWaves}.
The extension to a relativistic form is straightforward, and has been
used, in contexts related to the present work, to develop other algorithms
for plasma simulations%
  ~\cite{Evstatiev:2013:VarFormPtclSim,shadwick:14a,shadwick:14b}.
In Gaussian c.g.s. units, which we use throughout this work, it has the form
\begin{widetext}
\begin{equation}\label{eq:Low0}
  \mathpzc{L} = \int\!\ud\mathbf{x}_0 \ud\mathbf{v}_0
    \left[ - mc^2 \sqrt{1 - \left(\Dd{\mathbf{x}}{\tau}\right)^2}
           - q \phi(\mathbf{x}, t)
           + q\,\Dd{\mathbf{x}}{\tau} \cdot \mathbf{A}(\mathbf{x}, t)
    \right] \psi(\mathbf{x}_0, \mathbf{v}_0)\\
    +
    \frac{1}{8\pi}\int\!\ud\mathbf{x}
    \left[ \left( - \ptDd{\mathbf{A}}{\tau} - \grad\phi \right)^2
           - (\curl \mathbf{A})^2
    \right].
\end{equation}
\end{widetext}
Here the Boltzmann function
\(
  \psi(\mathbf{x}_0, \mathbf{v}_0)
\)
describes the phase-space distribution of particles at location
\(
  (\mathbf{x}_0, \mathbf{v}_0);
\)
$\phi$ and $\mathbf{A}$ denote the electromagnetic scalar and vector
potentials; and $\tau = ct$ denotes our independent variable.
The first integral describes the particles, including both their kinetic
energy and their interaction with the electromagnetic field.
The second integral describes oscillations of the electromagnetic field.

If we neglect the beam space charge, and consider only the
cavity eigenmodes, then $\phi = 0$. Furthermore, our beam has
\(
  | \ud\mathbf{x}_\perp / \ud\tau | \ll \ud{z} / \ud\tau \sim 1,
\)
while $|\mathbf{A}_\perp| \sim A_z$.
We therefore choose to neglect the transverse coupling terms.
We do not \emph{have} to do this: all the algorithms and computational
results described in this paper can be (and is some cases have been)
done without making this simplification; but doing so leaves us with
what we call the \emph{beam electromagnetic Low Lagrangian}:
\begin{widetext}
\begin{equation}\label{eq:Low1}
  \mathpzc{L}\left(
      \mathbf{x}, \Dt{\mathbf{x}}, \mathbf{A}, \ptDd{\mathbf{A}}{t}\right) =
    \int\!\ud\mathbf{x}_0 \ud\mathbf{v}_0 \left[
      - mc^2 \sqrt{1 - \left(\Dd{\mathbf{x}}{\tau}\right)^2}
             + q\,\Dd{z}{\tau} A_z(\mathbf{x}, t)
      \right] \psi(\mathbf{x}_0, \mathbf{v}_0)
      +
      \frac{1}{8\pi}\int\!\ud\mathbf{x}
      \left[ \left( \ptDd{\mathbf{A}}{\tau} \right)^2
             - (\curl\mathbf{A})^2
    \right].
\end{equation}
\end{widetext}
Our approximation---in essence neglecting $\mathbf{j}_\perp$---is made
in the interest of speed, as it dramatically reduces the number of required
source depositions, which dominates the computation time for self-consistent
simulations. For applications in which there is, for example, significant
gyrotron motion, one may need to retain the transverse coupling terms.

We assume the cavity eigenmodes are known and form a complete orthonormal basis,
so that we may decompose the vector potential in these modes:
\begin{equation}
  \mathbf{A} = \sum_\ell a_\ell(\tau) \mathbf{f}_\ell(\mathbf{x}),
\end{equation}
where the $\mathbf{f}_\ell$ denote spatial eigenmodes,
and $a_\ell$ the corresponding mode amplitudes.
Plugging this form into the Low Lagrangian, we obtain
\begin{widetext}
\begin{equation}\label{eq:Low2}
  \mathpzc{L} = \int\!\ud\mathbf{x}_0 \ud\mathbf{v}_0
    \left[ - mc^2 \sqrt{1 - \dot{\mathbf{x}}^2}
           + q\dot{z} \sum_\ell a_\ell\, \mathbf{z}\cdot\mathbf{f}_\ell(\mathbf{x})
    \right] \psi(\mathbf{x}_0, \mathbf{v}_0)
    +
    \frac{1}{8\pi}\sum_\ell
    \left[ \dot{a}_\ell^2 \int\!\ud\mathbf{x}\, |\mathbf{f}_\ell(\mathbf{x})|^2
           - a_\ell^2 \int\!\ud\mathbf{x}\, |\curl \mathbf{f}_\ell(\mathbf{x})|^2
    \right],
\end{equation}
\end{widetext}
where overdot denotes differentiation with respect to the proper time $\tau$.
In going from \eqref{Low1} to \eqref{Low2}, we have used the orthogonality
of electric and magnetic field eigenmodes to eliminate the cross-terms in
the second integral. As a convenience, we also define
the mode inductance
\(
  1/L_\ell = \frac{1}{4\pi} \int\!\ud\mathbf{x}\, |\curl \mathbf{f}_\ell|^2,
\)
and the mode capacitance
\(
  C_\ell = \frac{1}{4\pi} \int\!\ud\mathbf{x}\, |\mathbf{f}_\ell(\mathbf{x})|^2.
\)

To trace the particles, we introduce macroparticles by decomposing the
phase-space density in discrete shapes in the usual manner%
  ~\cite{shadwick:14a,shadwick:14b,webb_etal:2016c,webb:16b}:
\begin{equation}
  \psi(\mathbf{x}, \dot{\mathbf{x}}) = \sum_{j=1}^{N_\text{macro}}
      w_j \Lambda\left( \mathbf{x} - \mathbf{x}^{(j)} \right)
          \delta\left( \dot{\mathbf{x}} - \dot{\mathbf{x}}^{(j)} \right).
\end{equation}
Here $w_j$ denote the macroparticle weights,
$\delta$ the Dirac delta function,
and $\Lambda$ the normalized particle shape functions
(so that $\int\ud\mathbf{x}\, \Lambda = 1$).
This decomposition transforms the Lagrangian~\eqref{Low2} into a discrete set
of coupled macroparticle-electromagnetic-mode Lagrangians:
\begin{widetext}
\begin{equation}
  \mathpzc{L}\left(
      \mathbf{x}, \dot{\mathbf{x}}, a_\ell, \dot{a}_\ell
    \right) = \sum_{j=1}^{N_\text{macro}}
    \left[
      - w_j\, m c^2 \sqrt{1 - \bigl(\dot{\mathbf{x}}^{(j)} \bigr)^2}
      + w_j\, q \dot{z}^{(j)} \sum_\ell a_\ell\, F_\ell\bigl( \mathbf{x}^{(j)} \big)
    \right]
    + 
    \frac{1}{2} \sum_\ell
      \left[ C_\ell\, \dot{a}_\ell^2  - \frac{1}{L_\ell}\, a_\ell^2 \right],
\end{equation}
\end{widetext}
where
\begin{equation}
  F_\ell\bigl(\mathbf{x}^{(j)}\bigr) =
    \mathbf{z} \cdot \int\!\ud\mathbf{x}\,
      \mathbf{f}_\ell(\mathbf{x})\,
      \Lambda\bigl( \mathbf{x} - \mathbf{x}^{(j)} \bigr)
\end{equation}
represents the coupling of particle shape to field shape.
Each of the individual $\mathbf{x}^{(j)}$ and $a_\ell$ are dynamical variables
in this particle-in-mode discrete Lagrangian.

One approach to numerically integrating the resulting equations of motion
revolves around using time-discrete Lagrangians%
  ~\cite{marsden_west:01}.
This approach, however, typically yields an implicit algorithm,
which is substantially slower than an explicit version.
We therefore opt to develop explicit symplectic maps using a canonical formalism.

\subsection{Particle-Field Hamiltonian and a Split Operator Approach}

We can compute the Hamiltonian corresponding to the Lagrangian from the
canonical co\"ordinates and conjugate momenta for the particles and fields:
\begin{subequations}
\begin{equation}
  \mathbf{q}_j = \mathbf{x}_j,
\end{equation}
\begin{equation}
  \mathbf{p}^{(j)}_\perp = w_j\, mc \frac{\dot{\mathbf{x}}^{(j)}_\perp}%
                          {\sqrt{1 - \bigl(\dot{\mathbf{x}^{(j)}}\bigr)^2}},
\end{equation}
\begin{equation}
  p^{(j)}_z = w_j\, mc \frac{\dot{z}^{(j)}}%
                      {\sqrt{1 - \bigl(\dot{\mathbf{x}^{(j)}}\bigr)^2}}
                       - w_j\, q A_z,
\end{equation}
\begin{equation}
  Q_\ell = a_\ell,
\end{equation}
\begin{equation}
  P_\ell = C_\ell\, \dot{a}_\ell.
\end{equation}
\end{subequations}
The particle conjugate momenta are the usual,
but with the $\mathbf{A}_\perp$ components neglected.
The field conjugate momentum includes the mode capacitance,
which acts like an effective mass for the mode.

The Legendre transform to compute the coupled particle-field Hamiltonian is
\begin{equation}
  \mathcal{H} = \sum_j c\,\mathbf{p}_j \cdot \dot{\mathbf{q}}_j
              + \sum_\ell c\,P_\ell \dot{Q}_\ell - \mathpzc{L}.
\end{equation}
Carrying out this Legendre transform for these variables
yields the particle-field Hamiltonian
\begin{widetext}
\begin{equation}\label{eq:hamiltonian_unnormalized}
  \mathpzc{H} =
    \underbrace{
      \sum_j c\,\sqrt{
        \left( \mathbf{p}^{(j)}_\perp \right )^2
        + \left( p_z^{(j)} - w_j\, \frac{q}{c}
                     \sum_\ell Q_\ell F_\ell\left(\mathbf{q}^{(j)} \right) \right)^2
        + w_j^2 m^2 c^2}
    }_{\Hpc}
    \:+\,
    \underbrace{
      \frac{1}{2} \sum_{\ell}
      \left[
      \frac{P_\ell^2}{C_\ell} + \frac{1}{L_{\ell}} Q_\ell^2
      \right]
    }_{\Hf}.
\end{equation}
\end{widetext}
Here $\Hpc$ denotes the \emph{particle-coupling Hamiltonian}, which describes
both the particle dynamics and the particle-field coupling;
and $\Hf$ denotes the \emph{field Hamiltonian}, which describes the
harmonic oscillation of the independent field eigenmodes.

The mode quantities $C_\ell$ and $L_\ell$ are determined only up to within
a normalization constant. This means that the individual definitions of
$Q_\ell$ and $P_\ell$ will depend on the normalization convention used
for $\mathbf{f}_\ell$.
It will therefore prove useful to make a canonical transformation that combines
the two into a single scale-invariant quantity, which would be an intrinsic
quantity for a given cavity eigenmode.

By defining the canonically conjugate variables
\begin{subequations}
\begin{align}
  \Qell &= \sqrt{C_\ell}\, Q_\ell,\\
  \Pell &= \sqrt{\frac{1}{C_\ell}}\, P_\ell,
\end{align}
\end{subequations}
we transform the field and particle-coupling Hamiltonians into
\begin{subequations}
\begin{equation}\label{eq:Hfield}
  \Hf = \frac{1}{2} \sum_\ell \left[
      \Pell^2 + \frac{1}{L_\ell C_\ell} \Qell^2 \right]
\end{equation}
and
\begin{flalign}
  &\Hpc = c\, \sum_j \Biggl[
      \left( \mathbf{p}_\perp^{(j)} \right)^2 \notag \\
  & + \left( p_z^{(j)}
        - w_j\, \frac{q}{c} \sum_\ell \Qell \frac{1}{\sqrt{C_\ell}}
            F_\ell\left( \mathbf{q}^{(j)} \right) \right)^2
	+ w_j^2 m^2 c^2 \Biggr] ^{1/2}
\end{flalign}
\end{subequations}
All quantities in these Hamiltonians---including $\Qell$,
$\Pell$, and $F_\ell/\sqrt{C_\ell}$ -- are independent of our choice
of normalization for $\mathbf{f}_\ell$.
The invariant combination of the mode capacitance and inductance is
\begin{equation}
  \Omega_\ell^2 = \frac{1}{L_\ell C_\ell} =
    \frac{\int\!\ud\mathbf{x}\, |\curl\mathbf{f}_\ell(\mathbf{x})|^2}%
         {\int\!\ud\mathbf{x}\, |\mathbf{f}_\ell(\mathbf{x})|^2},
\end{equation}
which yields the eigenmode frequency, $\Omega_\ell$.

\subsection{Splitting the Hamiltonian}

The total particle-field Hamiltonian has no explicit time dependence,
so the symplectic map for going from time $\tau$ to time $\tau + h$
is simply
\begin{equation}\label{eq:MapHpcf}
  \mathscr{M}({\tau \rightarrow \tau + h})
      = \exp\left( -h\lieop{\Hpc + \Hf}\right).
\end{equation}
Here we use the colon notation introduced by Dragt%
  ~\cite{dragt_finn:1976,dragt:82,dragt_text:2016}
to denote that argument of the exponential is a Poisson-bracket Lie operator.
The full map \eqref{MapHpcf} is difficult to compute, but a symmetric splitting
yields a map that is both straightforward to compute and second-order accurate
in the step-size $h$:
\begin{subequations}\label{eq:full_map}
\begin{equation}
  \mathscr{M}(h) \approx \Mf(h/2) \Mpc(h) \Mf(h/2),
\end{equation}
where
\begin{align}
  \Mf(h/2) &= \exp\left( -\tfrac{h}{2} \lieop{\Hf} \right), \\
  \Mpc(h)  &= \exp\left( -h \lieop{\Hpc} \right),
\end{align}
\end{subequations}
and all maps are independent of the initial time $\tau$.
One can evaluate $\Mf$ exactly and $\Mpc$ to second order in $h$,
leading to an overall second-order accurate symplectic integrator
for both fields and particles.

\subsubsection{Field Map}

The field map $\Mf$, is simply the harmonic oscillator,
and it acts only on the field phase-space co\"ordinates:
\begin{multline}\label{eq:mapF}
  \begin{pmatrix} \Qell \\ \Pell \end{pmatrix}_\text{fin}
    = \Mf(\tfrac{h}{2})\circ
    \begin{pmatrix} \Qell \\ \Pell \end{pmatrix}_\text{ini}
    \\[1ex]
  = \begin{pmatrix}
       \cos \left(\Omega_\ell \frac{h}{2}\right)
           & \Omega_\ell^{-1} \sin\left(\Omega_\ell \frac{h}{2}\right) \\[1ex]
       -\Omega_\ell \sin\left(\Omega_\ell \frac{h}{2}\right)
           & \cos\left(\Omega_\ell \frac{h}{2}\right)
    \end{pmatrix}
    \begin{pmatrix} \Qell \\ \Pell \end{pmatrix}_\text{ini}
\end{multline}
As is usually the case with second-order splitting,
the two half-step field maps can, for simplicity, be combined
into a single full-step map---so long as one retains the half-step maps
at the \emph{ends} of the simulation.

\subsubsection{Particle-Coupling Map}

The particle-coupling map $\Mpc$ is not immediately integrable,
but it can be split using a method described by Wu, Forest, and Robin%
  ~\cite{wu_etal:03}
and exploited by Webb \emph{et al.}%
  ~\cite{webb_etal:2016c}
for a cylindrical electromagnetic algorithm.
This method involves tracking each particle with respect to its own proper time,
which allows us to re-write the Hamiltonian as an effectively
non-relativistic Hamiltonian given by
\begin{equation}\label{eq:NRHam}
  \Hpc =
      \sum_j \frac{
          \left( \mathbf{p}_\perp^{(j)} \right)^2
        + \left( p_z^{(j)}
            - w_j\, \frac{q}{c} \sum_\ell \Qell \frac{1}{\sqrt{C_\ell}}
                F_\ell \left( \mathbf{q}^{(j)} \right) \right)^2}%
      {2 w_j m \gamma^{(j)}},
\end{equation}
where $\gamma^{(j)}$ denotes the Lorentz factor of the $j^\text{th}$
macroparticle:\\
\begin{multline}\label{eq:gamma}
  \gamma^{(j)} w_j m c^2 = c\, \Biggl[
      \left( \mathbf{p}_\perp^{(j)} \right)^2 \\
   + \left(p_z^{(j)}
        - w_j\, \frac{q}{c} \sum_\ell \Qell \frac{1}{\sqrt{C_\ell}}
            F_\ell\left(\mathbf{q}^{(j)} \right) \right)^2
    + w_j^2 m^2 c^2 \Biggr] ^{1/2}
\end{multline}

Because the vector potential has no \emph{explicit} time dependence,
$\gamma^{(j)}$ is a constant of the motion for this Hamiltonian,
and the proper time of each macroparticle can be mapped to the lab time
simply by multiplying by $\gamma^{(j)}$.
Once written in the form \eqref{NRHam}, the Hamiltonian can be split
and integrated using a pair of half-drift maps for the transverse co\"ordinates,
and another map for the longitudinal co\"ordinate:
\begin{widetext}
\begin{equation}\label{eq:split_map_pc}
  \Mpc \approx
    \exp \left[ -\sum_j \frac{h}{2}
        \lieop[\Large]{
            \frac{\left( \mathbf{p}_\perp^{(j)} \right)^2}%
                 {2 w_j m \gamma^{(j)}}} \right]
    \exp \left[ -\sum_j h
        \lieop[\Large]{\frac{\left(p_z^{(j)}
        - w_j\, \frac{q}{c} \sum_\ell \Qell \frac{1}{\sqrt{C_\ell}}
            F_\ell\left(\mathbf{q}^{(j)} \right) \right)^2}%
                    {2 w_j m \gamma^{(j)}}} \right]
    \exp \left[ -\sum_j \frac{h}{2}
        \lieop[\Large]{
            \frac{\left( \mathbf{p}_\perp^{(j)} \right)^2}%
                 {2 w_j m \gamma^{(j)}}} \right].
\end{equation}
\end{widetext}
This result is again second-order accurate in $h$ and hence preserves
the overall order of the integrator.

The leading and trailing half-drifts are particularly simple:
\begin{multline}
    \exp \left[ -\sum_j \frac{h}{2}
        \lieop[\Large]{
            \frac{\left( \mathbf{p}_\perp^{(j)} \right)^2}%
                 {2 w_j m \gamma^{(j)}}} \right]
  \circ
  \begin{pmatrix} 
    \mathbf{p}_\perp^{(j)} \\[1ex]
    \mathbf{q}_\perp^{(j)}
  \end{pmatrix}
  = \\
  \begin{pmatrix} 
    \mathbf{p}_\perp^{(j)} \\[1ex]
    \mathbf{q}_\perp^{(j)} + h\, \mathbf{p}_\perp^{(j)}/ w_jm \gamma^{(j)}
  \end{pmatrix}.
\end{multline}

We can now apply the technique in~\cite{wu_etal:03}
to deal with the vector potential.
Lie transformations obey an important similarity transformation property:
\begin{equation}
  \lietr{f} \lietr{g} \lietr[-]{f} = \lietr{\lietr{f}g}.
\end{equation}
This property allows us to rewrite the central map in \eqref{split_map_pc}
as the product of three maps:
\begin{multline}
  \exp \left[ -\sum_j h
      \lieop[\Large]{%
          \frac{\left(p_z^{(j)} - w_j\,\frac{q}{c}
                    \sum_\ell \Qell \frac{1}{\sqrt{C_\ell}}
                              F_\ell\left(\mathbf{q}^{(j)}\right)
                \right)^2}%
               {2 w_j m \gamma^{(j)}}} \right] = \\
    \underbrace{
        \exp\left[ -\sum_j \sum_\ell w_j \frac{q}{c} \lieop[\Large]{
            \Qell \frac{1}{\sqrt{C_\ell}}
            \int\!\ud{z} F_\ell \left( \mathbf{q}^{(j)} \right)} \right]
    }_{\mathscr{A}_z} \times \\
    \underbrace{
        \exp\left[ -\sum_j h \lieop[\Large]{
            \frac{\left( p_z^{(j)} \right)^2}%
                 {2 w_j m \gamma^{(j)}}} \right]
    }_{\mathscr{D}_z} \times \\
    \underbrace{
        \exp\left[ \sum_j \sum_\ell w_j \frac{q}{c} \lieop[\Large]{
            \Qell \frac{1}{\sqrt{C_\ell}}
            \int\!\ud{z}\, F_\ell \left( \mathbf{q}^{(j)} \right)} \right]
    }_{\mathscr{A}_z^{-1}}.
\end{multline}
Moreover, each of these maps---the drift $\mathscr{D}_z$ and
the transformation $\mathscr{A}_z$---can be evaluated exactly:
\begin{subequations}\label{eq:mapsDzAz}
\begin{equation}
  \mathscr{D}_z \circ
      \begin{pmatrix} p_z^{(j)} \\ z^{(j)} \end{pmatrix}
    = \begin{pmatrix} p_z^{(j)} \\
               z^{(j)} + h \frac{p_z^{(j)}}{w_j m \gamma^{(j)}} \end{pmatrix},
\end{equation}
\begin{equation}
  \mathscr{A}_z \circ
      \begin{pmatrix} \Pell \\ \Qell \end{pmatrix} =
    \begin{pmatrix}
      \Pell
          + \sum_j w_j \frac{q}{c} (\sqrt{C_\ell})^{-1}
                       \int\!\ud{z}\, F_\ell\left(\mathbf{q}^{(j)}\right) \\
      \Qell
    \end{pmatrix}
\end{equation}
\begin{multline}
  \mathscr{A}_z \circ
      \begin{pmatrix} \mathbf{p}^{(j)} \\ \mathbf{q}^{(j)} \end{pmatrix} = \\
  \begin{pmatrix} 
    \mathbf{p}^{(j)} + \nabla_{\mathbf{q}}
        \sum_\ell w_j \frac{q}{c} (\sqrt{C_\ell})^{-1} \Qell
                      \int\!\ud{z}\, F_\ell\left(\mathbf{q}^{(j)}\right) \\
    \mathbf{q}^{(j)} \end{pmatrix}
\end{multline}
\end{subequations}

Because the algorithm is based on explicit symplectic maps, it is possible
to implement each of these maps as a single function, and the update sequence
is just a series of function calls. Once each of these maps is implemented,
it is straightforward to begin simulations.

Before going on to discuss our numerical results, we should clarify a point
that may seem mysterious: namely, how does the Lorentz $\gamma$ factor change?
The method given by Wu, Forest, and Robin is described in the context of
magnetostatic systems. Moreover, the absence of (explicit) time dependence in
the vector potential that appears in the Hamiltonian $\Hpc$ of \eqref{NRHam}
is consistent with a magnetostatic system, which cannot change the Lorentz
factor. And indeed the map $\Mpc$ derived from the Hamiltonian $\Hpc$ does
not do so.

The resolve the apparent mystery, recall that $\Hpc$ denotes just the
particle-coupling term in the full Hamiltonian: the remaining part is the
field Hamiltonian, given in~\eqref{Hfield}. When stepping through the full
map \eqref{full_map} to update the system for a single time step, one must
apply the field map $\Mf$ of \eqref{mapF}, which rotates the field
co\"ordinates, $\Qell$, and momenta, $\Pell$; and it is effectively
\emph{this} operation that updates---see \eqref{gamma}---the Lorentz $\gamma$
factor.

\section{Numerical Results}

To test the algorithm described in this paper, we considered a number of
configurations. In all tests, we considered a rectangular pillbox cavity
with a handful of test macroparticles.
We list the specific parameters in \tref{parameters}.
In all simulations, we considered four cavity modes: an accelerating mode,
two transverse dipole modes, and a single transverse quadrupole mode.

\begin{table}
\caption{Simulation parameters for testing the algorithm.}
\begin{ruledtabular}
\setlength\extrarowheight{2pt}
\begin{tabular}{ll}
  Parameter & Quantity \\
\hline
  Bunch Charge                         & \unit[192]{nC} \\
  \emph{e}-beam Energy                 & \unit[25.5]{MeV} \\
  Total Bunch Energy                   & \unit[$4.8 \times 10^7$]{ergs} \\
  Accelerating Gradient                & \unit[5.1]{MV/m} \\
  Cavity Length                        & \unit[10]{cm} \\
  Cavity Width $\times$ Cavity Height  & \unit[50]{cm} $\times$ \unit[40]{cm} \\
  Fundamental Frequency                & \unit[$2\pi\times 959$]{MHz}
\end{tabular}
\end{ruledtabular}
\label{tbl:parameters}
\end{table}

An artificially high bunch charge was selected to demonstrate the field energy
loss---otherwise the total energy of the bunch is so much smaller than
the total energy in the fields that the variations are not easy to discern.

In the first test, we considered a configuration with a single macroparticle
representing the entire bunch, accelerated from $\gamma = 50$ to $\gamma = 51$.
The particle is on-axis, so no higher order modes should be excited.
The result is shown in \fref{single_ptcl_on_axis}.
As can be seen, the energy conservation is excellent,
and none of the higher order modes are excited.
\begin{figure}[ht]
\includegraphics[width=0.48\textwidth]{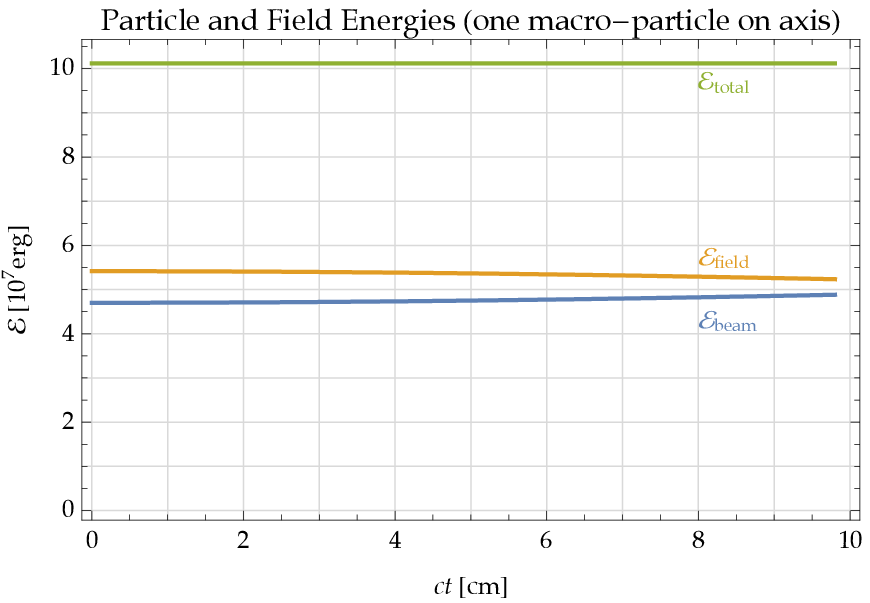}\\[2ex]
\includegraphics[width=0.48\textwidth]{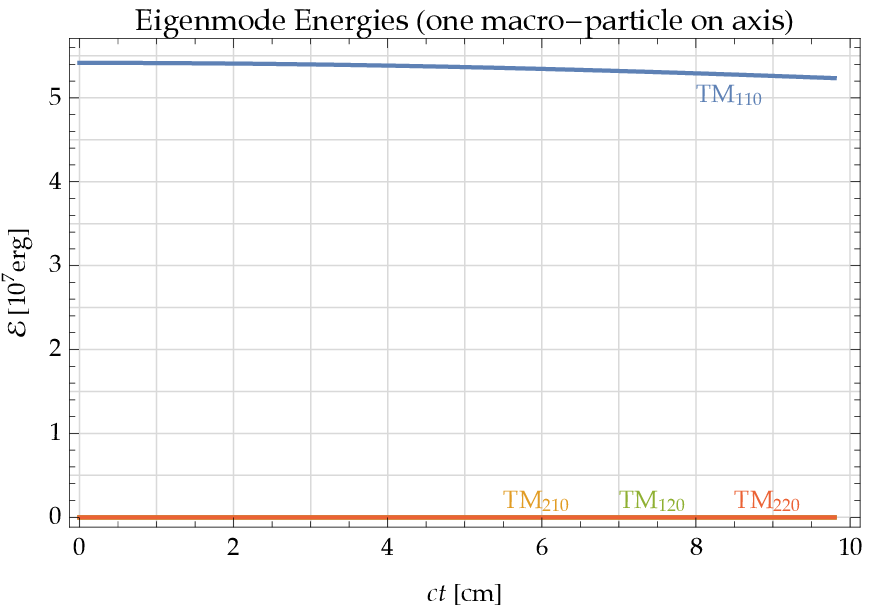}
\caption{Energy gain and eigenmode energy decomposition
  for the case of one macro-particle on axis.}
\label{fig:single_ptcl_on_axis}
\end{figure}

To test the excitation of higher order modes, we split the beam into
four macroparticles---one on-axis; one offset in $x$ and another offset in $y$
to excite dipole modes; and one offset along the $x$-$y$ diagonal to excite
both dipole and quadrupole modes.
\begin{figure}[ht]
\includegraphics[width=0.48\textwidth]{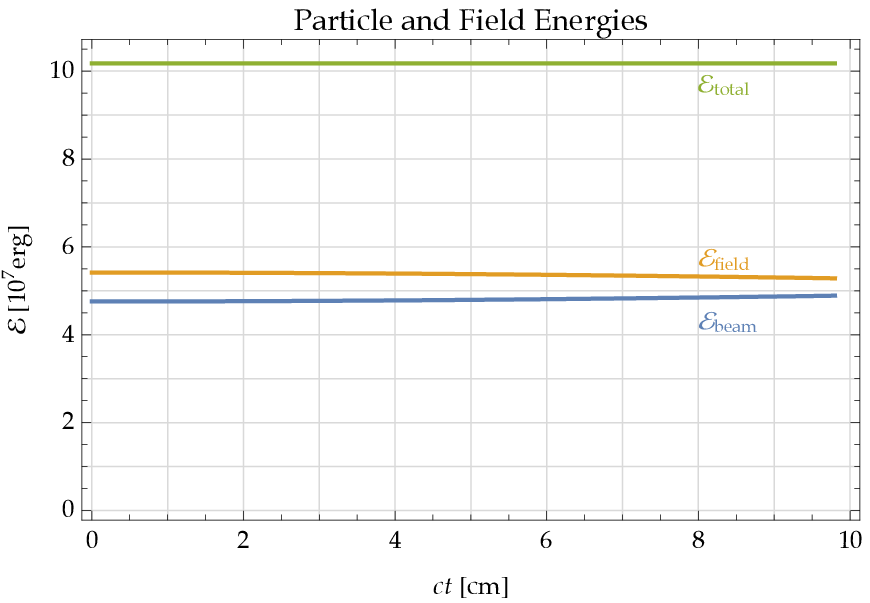}\\[2ex]
\includegraphics[width=0.48\textwidth]{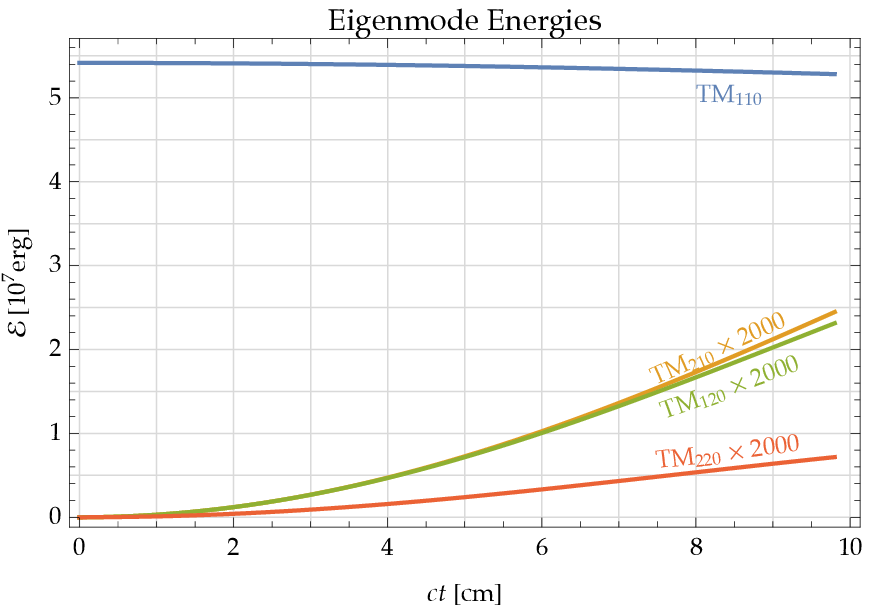}
\caption{Total particle-field energy and individual mode energies
  for the four-particle configuration.}
\label{fig:four_ptcl_short}
\end{figure}
As shown in \fref{four_ptcl_short}, the total energy is conserved to high
precision, and all the modes are excited, although the fundamental mode
sees the most variation.

Because this is a symplectic integrator, we expect the total energy to be
well-behaved for arbitrarily long time. To test this, we considered a much
longer rectangular pillbox, \unit[125]{cm} long, with the same transverse
eigenmodes. As shown in \fref{four_ptcl_long}, the algorithm retains stable
energy behavior for longer time scales.
\begin{figure}[ht]
\includegraphics[width=0.48\textwidth]{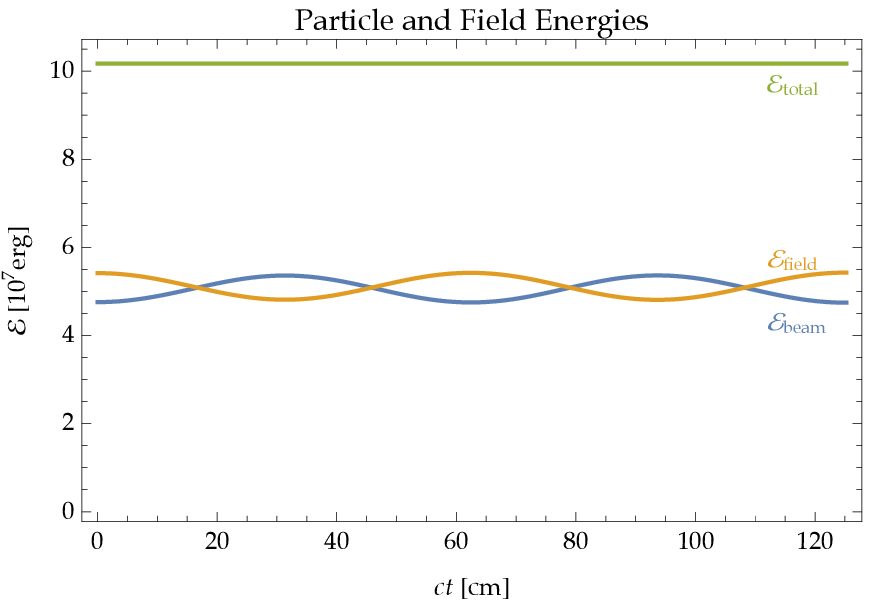}\\[2ex]
\includegraphics[width=0.48\textwidth]{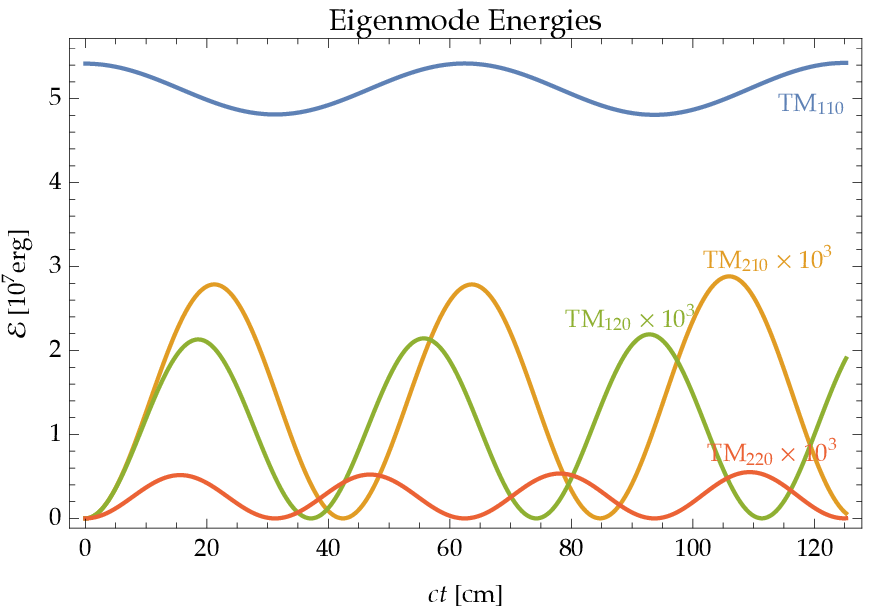}
\caption{Total particle-field energy and individual mode energies
  for the four-particle configuration in a longer cavity.}
\label{fig:four_ptcl_long}
\end{figure}
This captures oscillatory behavior as the beam accelerates and decelerates,
and the particles slip in phase for the higher order modes.

When simulating considerably longer cavities, \unit[$\sim$\,180000]{cm} 
in length, we detected a slow spurious decay in the total energy. 
Over this length, the error was $\sim$\,$10^{-6}\%$ total, as shown
in \fref{longterm_energy}. We have identified the origin of this spurious 
decay as numerical roundoff error in the field rotation matrix: for 
our implementation,the floating-point value of the determinant was 
very slightly less than~$1$. Over many time steps this can lead, 
depending on the roundoff error, to spurious cooling or heating of 
the system; but for a single bunch pass this is almost undetectable.
Furthermore, when advancing the fields in between bunch passes,
it suffices to perform a single matrix multiplication to advance
the eigenmodes.
Doing this reduces the roundoff-induced energy change in multi-bunch
and multi-pass simulations performed when studying, for example,
energy-recovery linacs.

\begin{figure}[ht]
\includegraphics[width=0.48\textwidth]{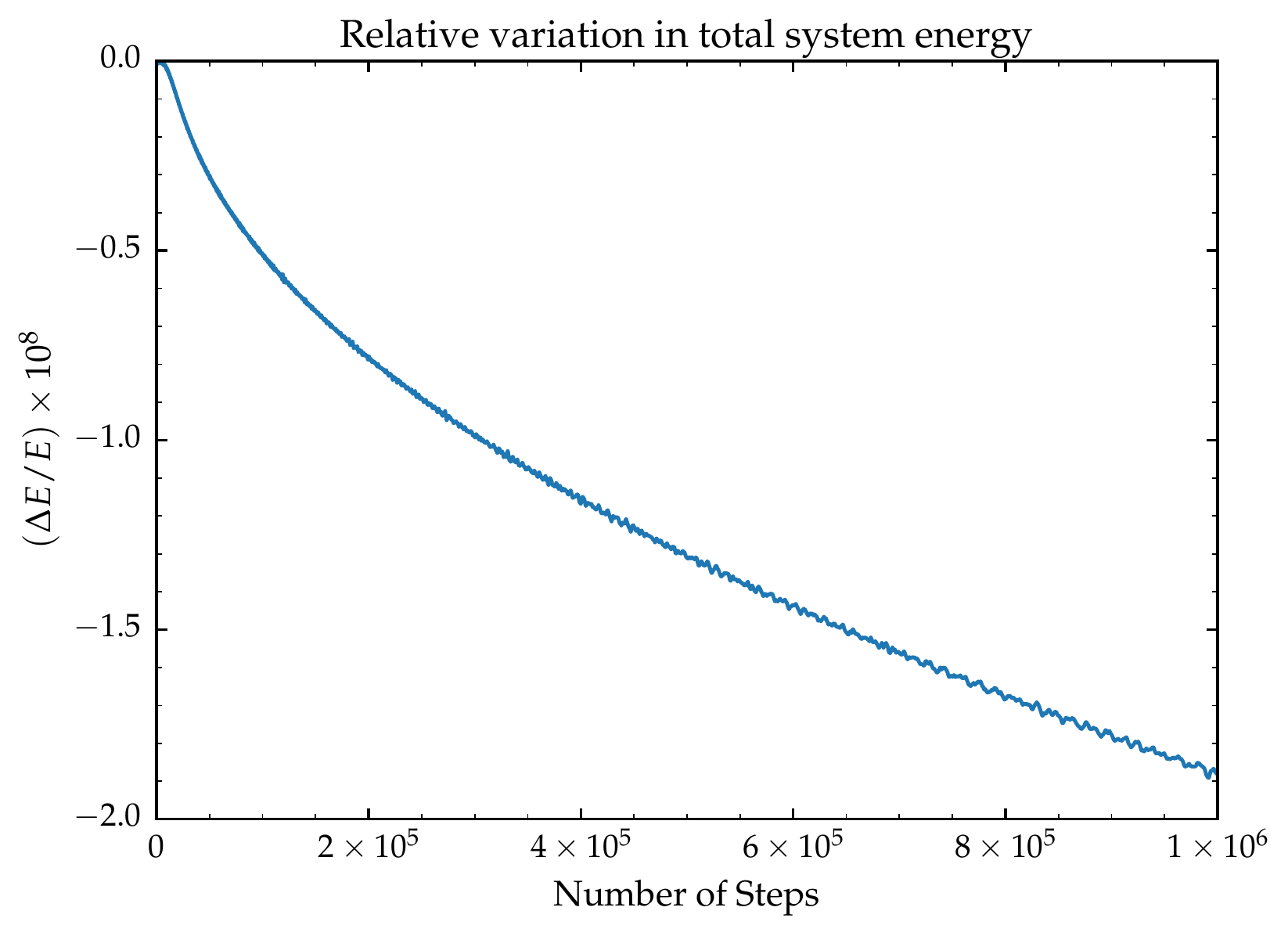}
\caption{Total system energy evaluated for 1000 particles and 6 modes in a
cavity of length  \unit[$\sim$\,1.8]{km}. 1 million steps of length
$\frac{\lambda}{30}$ produced relative energy ($\frac{\Delta E}{E}$)
losses on the order of $10^{-8}$.}
\label{fig:longterm_energy}
\end{figure}

\section{Algorithm Performance}


To demonstrate the second-order behavior of the algorithm, we performed
a simple test with a rectangular
\unit[10]{cm} $\times$ \unit[10]{cm} $\times$ \unit[20]{cm} 
cavity instantiated with three modes.
We propagated four particles through the cavity for a single step and
evaluated the change in the system Hamiltonian as the step size was varied.
The resulting behavior is shown in \fref{second_order_scaling}.
For decreasing step size, $h$, the error falls as $h^3$, indicative of a
second-order algorithm. For step sizes below \unit[1]{mm}, the single-step
error quickly approaches
machine precision, on the order of $10^{-14}$ relative error.

\begin{figure}[ht]
\includegraphics[width=0.48\textwidth]{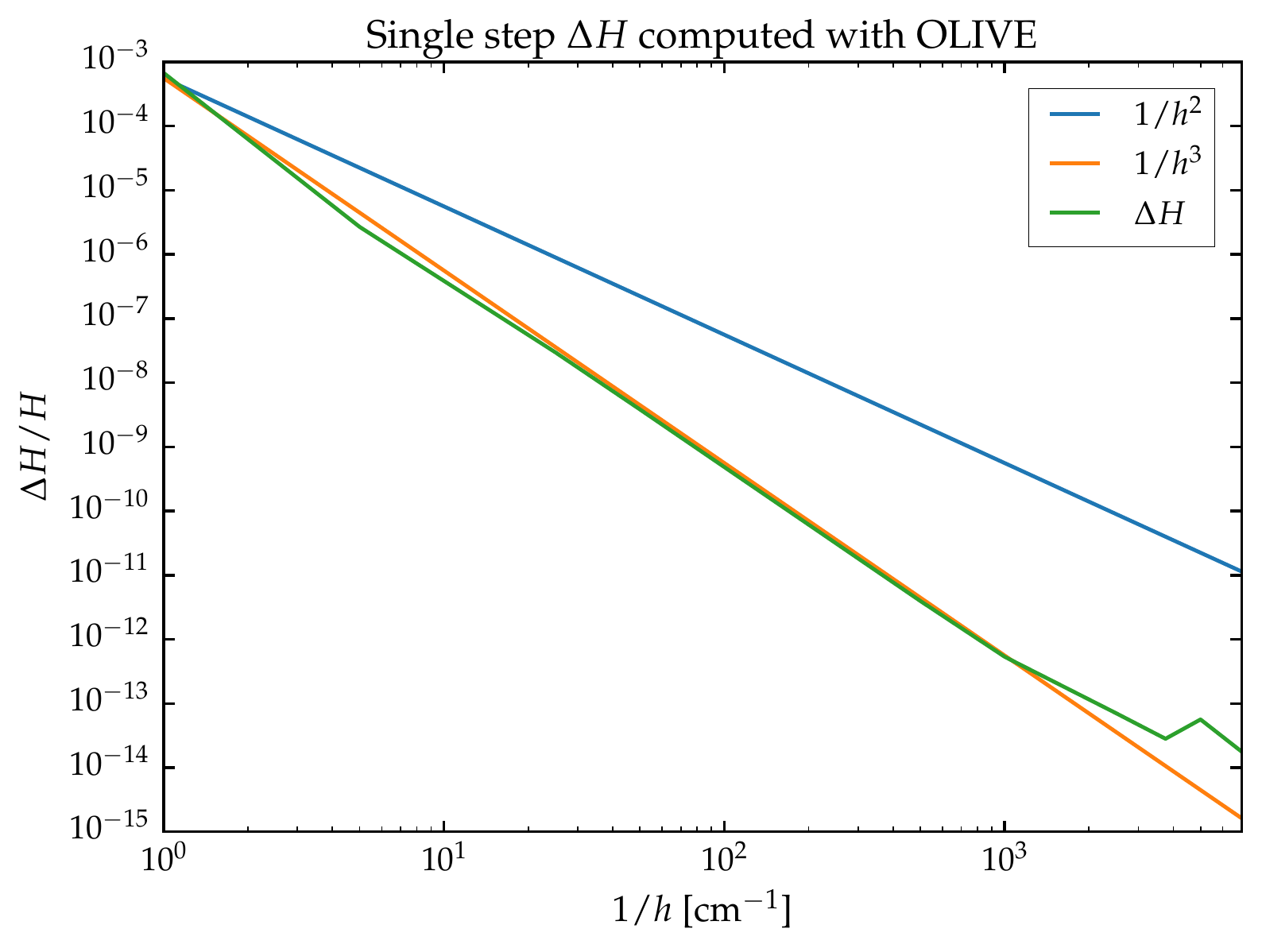}
\caption{The change in Hamiltonian evaluated after a single step
reveals 2nd order scaling with step size $h$.}
\label{fig:second_order_scaling}
\end{figure}

Initial benchmarks for the algorithm's performance demonstrate the scalability
to many hundreds of modes and thousands of particles. We performed 100~steps
of fixed step size on a single processor with varying macroparticle number and mode
number to illustrate the single-step evaluation time per-macroparticle-per-mode; the 
results are displayed in \fref{time_per_particle_per_mode}. Although Python-induced 
overhead associated with array creation hinders performance for small samples, the 
step times settle near $\Delta t \sim 0.5$\,$\mu$s per-particle-per-mode. When particle
and mode numbers cause array size to exceed cache size, a small drop in performance
occurs, as can be seen in to the top right of the figure.

\begin{figure}[ht]
\includegraphics[width=0.48\textwidth]{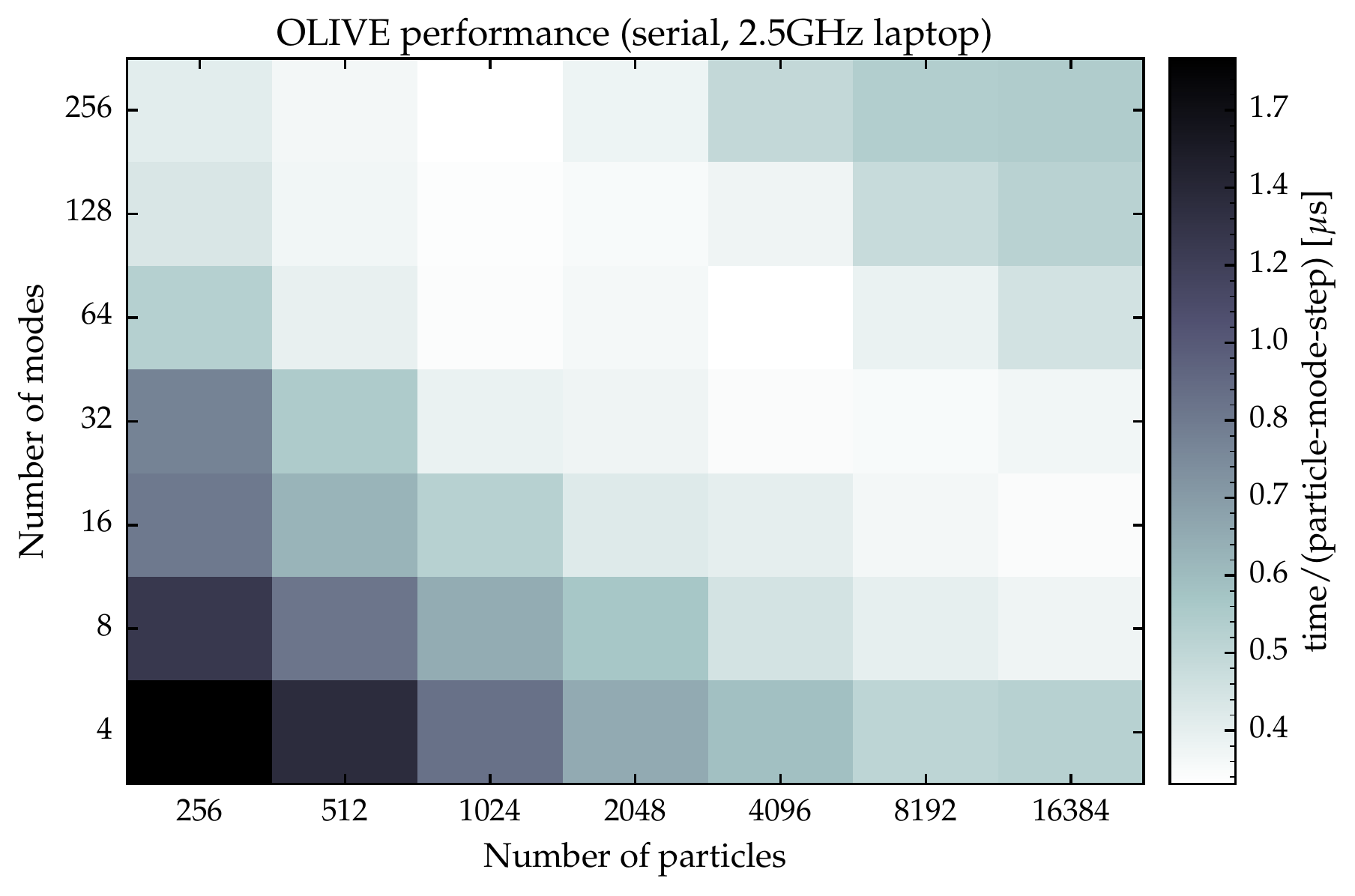}
\caption{Timings for a single step weighted per particle and per mode are
shown for upwards of 256~modes and 12800~particles. For large systems,
we find $\Delta t \sim 0.5$\,$\mu$s per-particle-per-mode.
}
\label{fig:time_per_particle_per_mode}
\end{figure}

\section{Conclusions}

We have presented a new approach to simulating beam loading in electromagnetic
cavities.  The approach is based on a particle-in-mode symplectic map approach
to electromagnetic charged-particle simulations.  The algorithm is suitable for
modeling a variety of vacuum electronic devices, including radiofrequency
cavities, traveling wave tubes, and klystrons.

For applications in energy recovery linacs, the beam remains relativistic at
all times, effectively suppressing space-charge forces. We therefore chose to
ignore it in our current implementation. But it \emph{is} possible to include
space-charge within the framework of our algorithm, and we mention two possible
approaches: One possibility is to stick with the Weyl gauge ($\phi = 0$) and
introduce additional high-order modes to represent the space-charge fields. When
the bunch is very short, the required modes will correspond to very-high-frequency,
essentially free-space, modes that are broadly separated from the cavity modes.
This approach will work well if the space-charge variation is predominantly
longitudinal, or if one includes the $\mathbf{j}_\perp$ terms we neglected in
transforming from the Lagrangian of \eqref{Low0} to that of \eqref{Low1}.
The other possibility is to use a spectral algorithm described in a paper
by one of the present authors%
  ~\cite{webb:16b}.

Another issue we did not address in our current implementation is the use
of realistic rf cavities. To model realistic cavity eigenmodes, one may use
the technology of \emph{generalized gradients} described in reference%
  ~\cite{abell:06}.
There the fields are computed on the basis of generalized gradients, which
one computes as integrals over a surface enclosing the relevant volume.
This surface integration acts to damp any imprecision in the simulated fields,
and it means that the field evaluations---including the derivatives---needed
for the maps in \eqref{mapsDzAz} can be computed to high accuracy.

The algorithm presented here avoids the various pitfalls of both the
finite-difference time-domain electromagnetic particle-in-cell and the
reduced-model approaches that are typically applied to this problem.

Because the algorithm is directly spectral, there are no grid aliasing
artifacts that can result in the numerical instabilities and dispersive errors
that tend to infect the full electromagnetic PIC simulations.
By using a modal decomposition, it is straightforward to identify,
directly from the dynamical data,
which cavity eigenmodes are excited and how they evolve---there is no need
for post-processing of the simulation data to extract the eigenmodes.
The approach also solves the load balancing problem, as the field data is
global, and the particles can be evenly distributed across multiple processors.
The approach is also much faster than the fully self-consistent approach:
our early simulations of the single particle traversing the pillbox cavity,
including the time to generate the figures, required less than a second using
an implementation that performed symbolic mathematical computations
and is far from optimized. See \fref{time_per_particle_per_mode} for
performance data based on a more sophisticated implementation.

The symplectic particle-in-mode (SymPIM) algorithm described in this
paper in many ways resembles the reduced models, but it is distinct
in a number of important ways.
Because the cavity is not treated as a thin element, it is possible
to model the complex phase-space of realistic beams, such as those
with large energy spreads, or those disrupted by beam-beam collisions
or various wakefield instabilities.
The SymPIM algorithm thus lies between the fully self-consistent
approach---which includes all resolvable cavity modes at great
computational expense---and the reduced-model approach---which neglects
many details of the beam distribution and how it evolves within the cavity.
This makes the SymPIM algorithm suitable for determining the validity
of reduced models, as well as for modeling those cases where the reduced
models do not apply.

\section{Acknowledgements}

This material is based upon work supported by the U.S. Department of Energy,
Office of Science, Office of Nuclear Physics under Award Number~{DE-SC0015211}.

\providecommand{\noopsort}[1]{}

\end{document}